# Active Power Thermal Feasibility Assessment of EV Integration with DER Scheduling and Distribution Network Reconfiguration

Rida Fatima
*Student Member, IEEE*
Department of Electrical and Computer Engineering
University of Houston
Houston, TX, USA
rfatima4@uh.edu

Linhan Fang
*Student Member, IEEE*
Department of Electrical and Computer Engineering
University of Houston
Houston, TX, USA
lfang7@uh.edu

Xingpeng Li
*Senior Member, IEEE*
Department of Electrical and Computer Engineering
University of Houston
Houston, TX, USA
xli82@uh.edu

***Abstract*— The rapid growth of electric vehicle (EV) adoption poses operational and economic challenges for distribution systems, including increased line loading and network congestion. As a practical near-term alternative to infrastructure reinforcement, distribution network topology reconfiguration (DNTR) can redistribute power flows, reduce operating costs, and improve operational flexibility. This paper presents a data driven operational feasibility framework for evaluating EV integration under increasing demand levels across four configurations: standard distribution network (SDN), SDN with DNTR (SDNTR), SDN with distributed energy resources (SDN-DER), and SDNTR with DERs (SDNTR-DER). Smart meter derived EV charging behavior is used to construct nodal EV demand profiles, which are incorporated into a mixed integer linear programming based active power screening model for a modified IEEE 33 bus distribution system. Feasibility is assessed with respect to nodal active power balance, line thermal limits, radiality and connectivity, DER limits, and BESS intertemporal operation. A bisection search identifies maximum active power thermal feasible EV demand levels of 51%, 68%, 85%, and 102% for SDN, SDN-DER, SDNTR, and SDNTR-DER, respectively. The 102% result represents full residential EV adoption with a 2% additional charging demand margin. At 0% EV penetration, SDNTR-DER reduces operating cost by 54% relative to SDN, representing the baseline economic value of coordinated DER scheduling and DNTR. The reported limits represent active power thermal feasibility boundaries rather than full AC hosting capacity limits.**

## NOMENCLATURE

*Sets:*

| | |
|---|---|
| $S$ | Set of substations |
| $N$ | Set of buses |
| $K$ | Set of lines |
| $K(n-)$ | Set of lines where bus $n$ is the from-bus |
| $K(n+)$ | Set of lines where bus $n$ is the to-bus |
| $S(n)$ | Substation at bus $n$ |
| $G$ | Set of generators |
| $P$ | Set of PV units |
| $B$ | Set of battery energy storage systems (BESS) |
| $T$ | Set of time periods (hours in a day) |

*Parameters:*

| | |
|---|---|
| $C_{s,t}^{sub}$ | Electricity price at substation $s$ at time $t$ |
| $C_{g,t}^{gen}$ | Generation cost for generator $g$ at time $t$ |
| $soc_{min}$ | Minimum state of charge |
| $soc_{max}$ | Maximum state of charge |
| $T_{chg}$ | Charging Duration |
| $T_{dchg}$ | Discharging Duration |
| $\eta_b^{chg}$ | BESS charging efficiency |
| $\eta_b^{dchg}$ | BESS discharging efficiency |
| $D_{n,t}$ | Load at bus $n$ at time $t$ |
| $x_k$ | Reactance of line $k$ |
| $M$ | Big-M constant |
| $Rating_k$ | Thermal limit of line $k$ |
| $P_{p,t}^{pv}$ | PV generation from unit $p$ at time $t$ |
| $P_g^{max}$ | Maximum power produced by $g$ |
| $P_g^{min}$ | Minimum power produced by $g$ |
| $E_b^{cap}$ | Energy capacity of BESS $b$ |
| $D_{n,t}^{EV}$ | EV demand at bus $n$ and time $t$ |
| $E_b^{init}$ | Initial stored energy in BESS $b$ |

*Variables:*

| | |
|---|---|
| $\theta_{n,t}$ | Phase angle of bus $n$ at time $t$ |
| $J_{k,t}$ | Binary status of branch $k$ at time $t$ |
| $P_{s,t}^{sub}$ | Power purchased from substation $s$ at time $t$ |
| $P_{k,t}^{line}$ | Power flow on line $k$ at time $t$ |
| $P_{p,t}^{curt}$ | PV curtailed power from unit $p$ at time $t$ |
| $P_{b,t}^{dchg}$ | Discharging power of BESS $b$ at time $t$ |
| $P_{b,t}^{chg}$ | Charging power of BESS $b$ at time $t$ |
| $P_{g,t}^{gen}$ | Power produced by $g$ at time t |
| $c_{b,t}$ | Binary charging status of BESS $b$ |
| $d_{b,t}$ | Binary discharging status of BESS $b$ |
| $E_{b,t}$ | Stored energy in BESS $b$ at time $t$ |
| $E_{b,24}$ | Stored energy in BESS $b$ *at hour t=24* |
| $F_{k,t}^{+}$ | Artificial connectivity flow on line $k$ in the forward direction. |
| $F_{k,t}^{-}$ | Artificial connectivity flow on line $k$ in the reverse direction. |

## I. INTRODUCTION

The rapid adoption of electric vehicles (EVs) is introducing new operational challenges for distribution networks that were not originally designed to handle such dynamic and concentrated charging demand [1]. EV charging loads can be substantial, often comparable to household peak demand, and are frequently temporally correlated as many users charge during evening or nighttime periods. This clustering can create sharp load spikes, stress network components, and reduce operational margins [2]. A key challenge from EV integration is line congestion. Distribution feeders have limited thermal capacity, and concentrated charging can cause certain lines to operate near or above their ratings, potentially requiring costly dispatch actions or load curtailment. Unlike transmission networks, distribution systems often have fewer redundancies, making congestion relief more constrained [3]. As EV adoption increases, these limitations can emerge at specific locations and hours rather than uniformly across the feeder. This makes the spatial allocation and temporal pattern of charging important when assessing how much additional EV demand the existing network can accommodate.

High charging demand may also create voltage and loss related concerns, although this paper focuses on active power congestion, line capacity feasibility, and operating cost. EV integration can increase reliance on expensive supply sources when network constraints limit the use of local generation or renewable resources. This is especially relevant when distributed energy resources (DERs), including photovoltaic (PV) units, natural gas generators (NGs), and battery energy storage systems (BESS), cannot be fully utilized because of congestion [4]-[7]. Coordinated DER operation and distribution network topology reconfiguration can redirect power flows, relieve constrained corridors, reduce operating costs, improve network flexibility, and potentially defer infrastructure upgrades.

## II. Literature Review

High penetrations of PVs and EVs can intensify feeder overloading, renewable curtailment, congestion, increased losses, equipment overloading, and degraded service quality [8]-[12]. At the same time, EVs have been recognized as enablers for integrating intermittent renewables through coordinated operation with DERs [13]-[16].

To address congestion and other operational issues, DNTR has emerged as an effective strategy by optimally controlling sectionalizing and tie-line switches to alter network topology [17]-[18]. Literature reports a range of DNTR methods, including classical loss reduction and load balancing approaches [19], [20], heuristic and active network reconfiguration methods [21], and PSO-based techniques for loss reduction and distributed generation integration [22]-[24]. These studies demonstrate that DNTR is not limited to congestion mitigation but is also a versatile tool for addressing multiple operational challenges in modern distribution networks. Building on these insights, the present work applies DNTR within an operational cost and EV operational feasibility assessment framework to manage congestion, enhance economic efficiency, and evaluate active power thermal feasibility under increasing EV demand in networks with DER integration.

### *A. Research Gaps & Contributions*

Despite progress in distribution system operation with EVs, DERs, and topology reconfiguration, most existing studies emphasize either EV charging coordination, DER scheduling, EVCS planning, or loss-oriented reconfiguration. For example, reconfiguration studies such as [19]-[24] primarily target loss reduction, load balancing, or DG integration, while EV focused works [1], [2], [6] emphasize charging coordination and scheduling. However, the marginal and joint operational value of DNTR and DER integration under a common data-driven EV demand model remains insufficiently quantified in the existing literature. Fewer studies provide a controlled operational assessment that isolates and quantifies the individual and joint contributions of DER integration and DNTR under data-driven EV demand growth. This paper addresses this gap by developing a data informed MILP-based operational screening framework that evaluates active power feasibility, congestion relief, and operating cost across standard and reconfigurable feeder configurations. The main contributions are:

i. A smart meter informed EV charging load mapping process that incorporates statistically generated charging profiles into nodal EV demand scenarios under increasing penetration levels.
ii. A MILP-based active power operational screening model that co-optimizes substation imports, DER dispatch, BESS operation, and DNTR while enforcing line thermal limits, radiality, and substation connected topology.
iii. A bisection-based characterization of the active power thermal feasibility boundary across the four configurations, identifying the EV demand scaling at which each configuration transitions from feasible to infeasible and quantifying the corresponding thermal upgrade-deferral headroom for the studied feeder.

## III. Mathematical Modeling

The MILP model co-optimizes substation imports, DER dispatch, BESS operation, and topology reconfiguration over a 24-hour horizon for four configurations: SDN, SDNTR, SDN-DER, and SDNTR-DER. A case is deemed operationally feasible if nodal active power balance, line thermal limits, radiality/connectivity, DER limits, and BESS intertemporal constraints are satisfied over the full horizon. The objective function (1) minimizes total operating cost.

$$min\{\sum_{s\in S}\sum_{t\in T} C_{s,t}^{sub} P_{s,t}^{sub} + \sum_{g\in G}\sum_{t\in T} C_{g,t}^{gen} P_{g,t}^{gen}\} \tag{1}$$

The first term represents imported electricity cost at substation $s$ and time $t$ (time varying price × imported power); the second term is the local generation cost from generator $g$ at time $t$. A linearized active power network model preserves MILP tractability. For each line $k = (i, j)$, buses $i$ and $j$ denote the from and to bus endpoints. The governing constraints are as follows:

$$\begin{aligned}\sum_{k\in K(n+)} P_{k,t}^{line} - \sum_{k\in K(n-)} P_{k,t}^{line} + \sum_{s\in S(n)} P_{s,t}^{sub} \\ + \sum_{p\in P(n)} (P_{p,t}^{PV} - P_{p,t}^{curt}) - \sum_{b\in B(n)} P_{b,t}^{chg} \\ + \sum_{b\in B(n)} P_{b,t}^{dchg} + \sum_{g\in G(n)} P_{g,t}^{gen} \\ = D_{n,t} + D_{n,t}^{EV}\end{aligned} \tag{2}$$

$$-Rating_k J_{k,t} \le P_{k,t}^{line} \le Rating_k J_{k,t} \tag{3}$$

$$-M(1 - J_{k,t}) \le P_{k,t}^{line} - \frac{\theta_{i,t} - \theta_{j,t}}{x_k} \le M(1 - J_{k,t}) \tag{4}$$

$$\sum_{k\in K} J_{k,t} = |N| - |S| \tag{5}$$

$$0 \leq F_{k,t}^{+} \leq (|N| - |S|) J_{k,t} \tag{6}$$

$$0 \leq F_{k,t}^{-} \leq (|N| - |S|) J_{k,t} \tag{7}$$

$$\sum_{k \in K(n+)} F_{k,t}^{+} + \sum_{k \in K(n-)} F_{k,t}^{-} - \sum_{k \in K(n-)} F_{k,t}^{+} - \sum_{k \in K(n+)} F_{k,t}^{-} = 1 \tag{8}$$

$$P_{g}^{min} \leq P_{g,t}^{gen} \leq P_{g}^{max} \tag{9}$$

$$0 \leq P_{p,t}^{curt} \leq P_{p,t}^{PV} \tag{10}$$

$$soc_{min} * E_{b}^{cap} \leq E_{b,t} \leq soc_{max} * E_{b}^{cap} \tag{11}$$

$$E_{b}^{init} = E_{b,24} \tag{12}$$

$$c_{b,t} + d_{b,t} \leq 1 \tag{13}$$

$$0 \leq P_{b,t}^{chg} \leq \frac{E_{b}^{cap}}{T_{chg}} * c_{b,t} \tag{14}$$

$$0 \leq P_{b,t}^{dchg} \leq \frac{E_{b}^{cap}}{T_{dchg}} * d_{b,t} \tag{15}$$

$$E_{b,1} = E_{b}^{init} + \eta_{b}^{chg} P_{b,1}^{chg} - \frac{P_{b,1}^{dchg}}{\eta_{b}^{dchg}} \tag{16}$$

$$E_{b,t} = E_{b,t-1} + \eta_{b}^{chg} P_{b,t}^{chg} - \frac{P_{b,t}^{dchg}}{\eta_{b}^{dchg}} \tag{17}$$

Constraint (2) enforces nodal active power balance among line flows, substation imports, PV, BESS, DG, base load, and EV demand, including PV curtailment and BESS charging. Constraint (3) limits line flows according to branch status, while (4) applies the DC power flow relation only to closed lines. Constraint (5) sets the number of closed branches required for radial operation, and (6)-(8) use artificial connectivity flows to ensure that every non substation bus remains connected to a substation without isolated islands. Constraint (9) bounds DG output, while (10) limits PV curtailment to available generation. Constraint (11) enforces BESS SOC limits, and (12) requires the final stored energy to equal the initial value. Constraint (13) prevents simultaneous charging and discharging, (14)-(15) limit charging and discharging power, and (16)-(17) track stored energy considering charging and discharging efficiencies.

A linearized active power network model is adopted to maintain MILP tractability for the time coupled DNTR problem with binary switching, DER dispatch, and BESS intertemporal constraints [3], [17]. The same active power screening assumptions are applied to all four configurations to enable a controlled comparison of cost and feasibility trends. The framework is therefore intended as an active power thermal screening tool that identifies thermally binding EV demand levels before detailed AC studies. The reported limits should be interpreted as active power thermal feasibility boundaries rather than full AC hosting capacity limits because voltage magnitude, reactive power, losses, and other AC network constraints are not modeled.

## IV. Test Case Description

The test system is based on the IEEE 33 bus radial distribution network in [19], with a nominal voltage of 12.66 kV, and is modified to include two substations, DERs, and controllable branches. It consists of 33 buses, two substations, multiple DERs, and controllable branches for reconfiguration. The base topology is shown in Fig. 1. A 24-hour scheduling horizon is used for simulations. Substations are located at buses 1 and 33, while PV units are placed at buses 15, 16, 21, and 27. Two natural gas DGs are installed at buses 23 and 24, and four BESS units are co-located with PV buses for load shifting, peak shaving, and renewable integration.

In the DNTR enabled cases, the statuses of existing branches are optimized to enable topology reconfiguration for congestion relief and operating cost reduction; since these branches are assumed to already exist, no capital cost is considered. The system is modified with respect to DG placement and PV/BESS additions to resemble a realistic feeder with high renewable penetration.

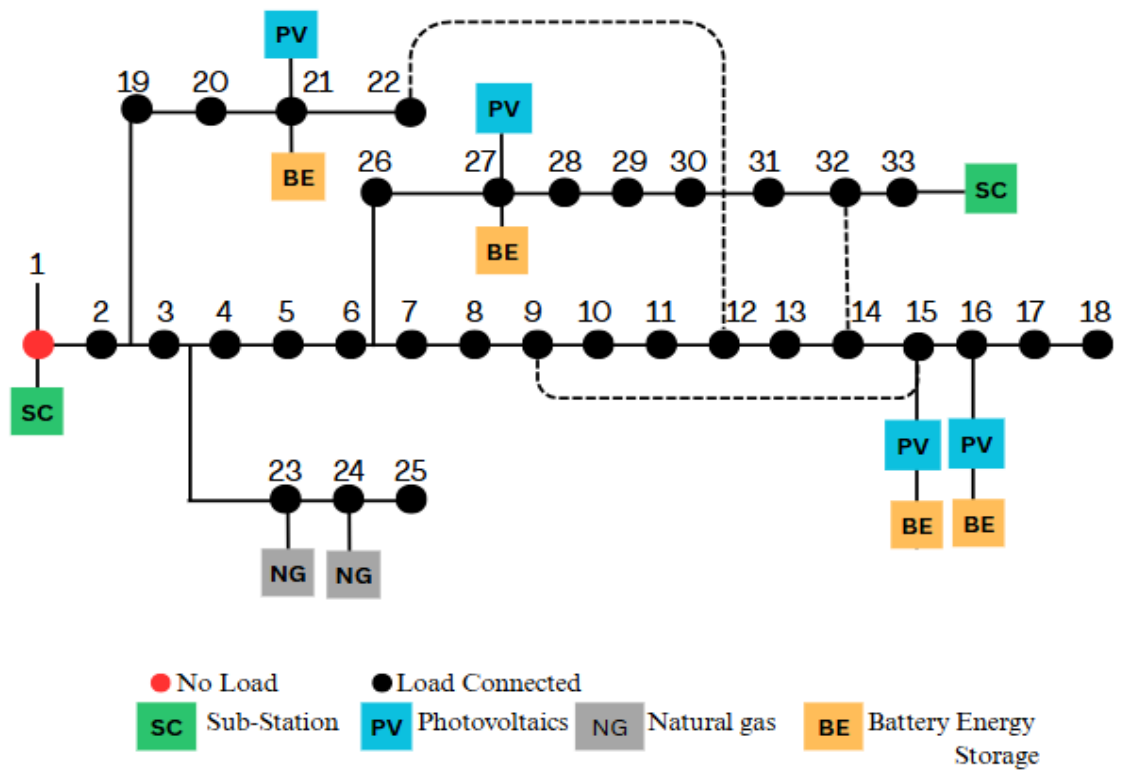

Figure 1. Modified IEEE 33-bus system.

This configuration creates diverse operational scenarios requiring coordinated substation imports, DG dispatch, renewable utilization, and BESS operation to minimize costs while satisfying all constraints.

## V. Electric Vehicle Integration

The 15-minute EV charging scenarios are aggregated into hourly nodal profiles, and the expected profile is used as $D_{n,t}^{EV}$ in the 24-hour MILP. Charging events are filtered and characterized by energy, duration, and start and end times using kernel density estimation. These distributions are used in Monte Carlo simulation to generate 1,200 annual scenarios with a 90% daily charging probability, including low, normal, and high-power charging profiles. The resulting scenarios provide representative EV demand inputs, while stochastic or robust optimization over individual scenarios is outside the scope of this screening study. To assess sensitivity to coincident high charging demand, the actual scenario with a daily aggregate peak closest to the 95th percentile is selected and its complete 24-hour nodal profile is evaluated using the same network model and feasibility procedure as the expected profile case.

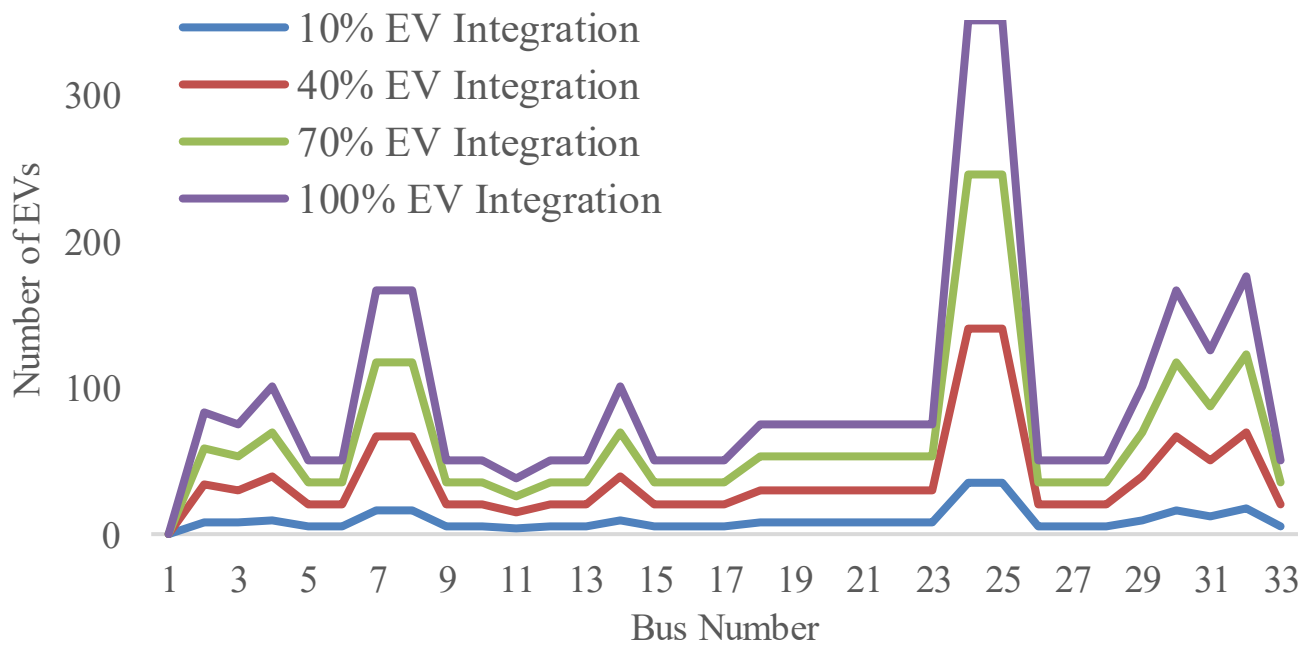

Figure 2. EV allocation across buses for the studied penetration levels.

The 100% EV case represents one EV per residential home, while the 10%, 40%, and 70% cases are obtained by proportionally scaling the full adoption charging profile and bus allocation shown in Fig. 2. Values above 100% represent additional scaling of charging demand rather than EV adoption above 100%. For example, 102% corresponds to 1.02 times the one EV per home charging demand.

Fig. 3 compares the normalized temporal shapes of the base load and EV charging load profiles over the 24-hour horizon [25]. The EV load is highly time dependent, with pronounced evening and nighttime peaks around hours 19-23, reflecting typical charging behavior after users return home. This temporal and spatial variability significantly affects network loading, feasibility, and operational costs.

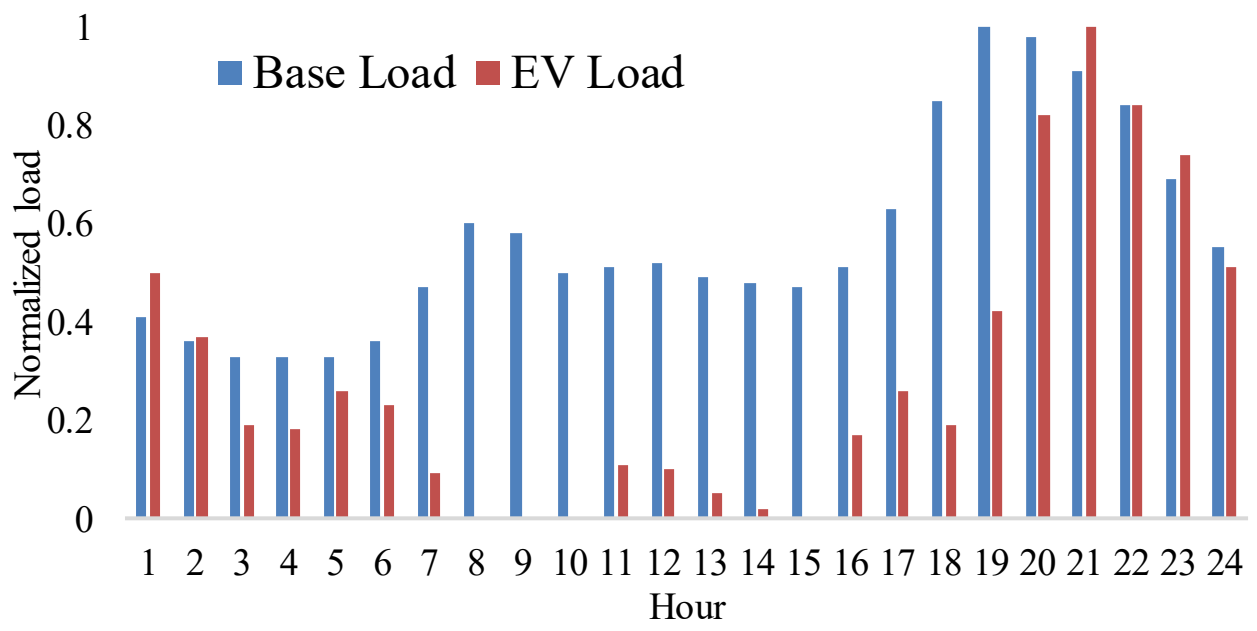

Figure 3. Normalized hourly baseload and EV charging load profiles.

## VI. Results and Analysis

The MILP model was implemented in Pyomo and solved using Gurobi 11.0.3 with a 1% MIP gap, with each 24-hour case solving in under one minute. Fig. 4 compares operating costs, while Table I summarizes cost and feasibility up to 100% EV penetration. Relative to SDN, SDNTR-DER reduces operating cost by 54%, 47%, and 38% at 0%, 10%, and 40% EV penetration, respectively; the 54% reduction at 0% reflects the baseline value of coordinated DER scheduling and DNTR. The active power thermal feasibility limits are 51%, 68%, 85%, and 102% for SDN, SDN-DER, SDNTR, and SDNTR-DER, respectively. At high EV demand, infeasibility is driven by evening peak line flow violations. DNTR relieves congestion by redistributing imports across substation paths, while DERs and BESS reduce loading on constrained feeder corridors.

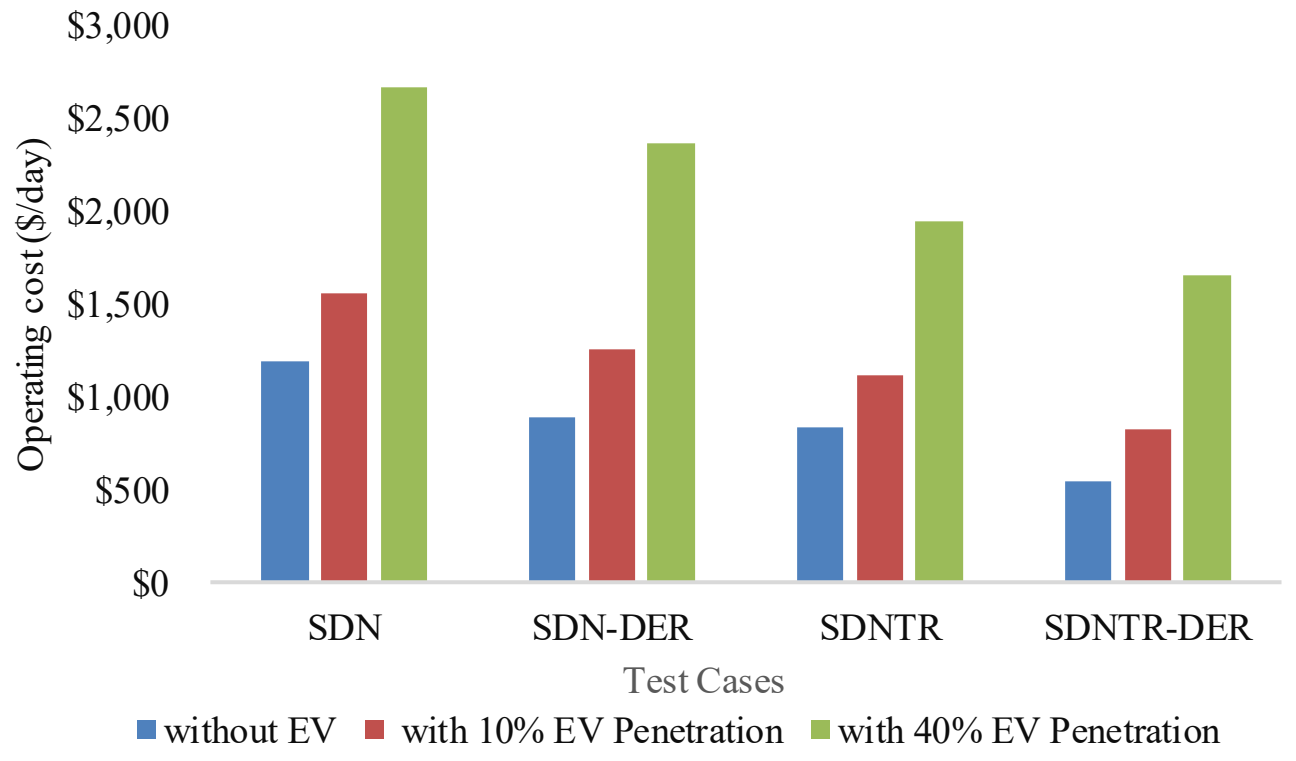

Figure 4. Operating cost comparison for EV penetration levels where all four configurations are feasible.

Table I shows that DER integration provides a substantial operating cost benefit at low to moderate EV penetration, while DNTR provides the largest extension of the active power thermal feasibility boundary. The two effects combine constructively: SDNTR-DER reduces operating cost by 54%, 47%, and 38% at 0%, 10%, and 40% EV penetration, respectively, while extending the active power thermal feasible EV demand boundary from 51% to 102% of the full adoption charging profile. Physically, DERs displace substation imports at low-cost hours while DNTR relieves the corridors that constrain evening peak imports.

Table I. Operating Cost and Feasibility Under Various EV Penetration Levels

| | No EV | 10% EV Penetration | 40% EV Penetration | 70% EV Penetration | 100% EV Penetration |
|---|---|---|---|---|---|
| **SDN** | $1,190 | $1,559 | $2,667 | Infeasible | Infeasible |
| **SDN-DER** | $883 | $1,251 | $2,357 | Infeasible | Infeasible |
| **SDNTR** | $832 | $1,108 | $1,940 | $2,921 | Infeasible |
| **SDNTR-DER** | $545 | $824 | $1,649 | $2,582 | $3,652 |

Table II reports the status of Line 1 in SDNTR-DER, connecting Substation 1 to the network, during hours 20-24 (the evening EV peak and high-cost period of Substation 1). At lower EV penetration, Line 1 is opened during selected hours to shift import to the lower cost supply path; as EV demand grows, it remains closed more frequently at 70% and 100% penetration to provide additional import capability and maintain feasibility.

These switching decisions show how DNTR redistributes power flows during peak EV demand to maintain feasibility under the same line thermal limits, with only limited switching activity across all feasible cases.

TABLE II. LINE 1 SWITCHING STATUS FOR SDNTR-DER DURING EVENING PEAK HOURS

| Hour | No EV | 10% EV Penetration | 40% EV Penetration | 70% EV Penetration | 100% EV Penetration |
|---|---|---|---|---|---|
| **20** | 1 | 1 | 1 | 1 | 1 |
| **21** | 1 | 1 | 1 | 1 | 1 |
| **22** | 0 | 0 | 0 | 1 | 1 |
| **23** | 0 | 0 | 0 | 0 | 1 |
| **24** | 0 | 0 | 0 | 0 | 1 |

As EV demand increases, DNTR shifts from supporting economic dispatch to preserving import capability during the evening charging peak. DER and BESS complement this by reducing active power transfer through constrained feeder corridors.

A bisection search identifies the maximum active power thermal feasible EV demand for each configuration and stops when the interval between the last feasible and first infeasible levels is below 1 percentage point. The resulting boundaries are 51% for SDN, 68% for SDN-DER, 85% for SDNTR, and 102% for SDNTR-DER, where 102% represents the full adoption profile plus a 2% uniform demand margin.

Relative to SDN, DER and DNTR increase the feasibility boundary by 17 and 34 percentage points, respectively, while their coordinated use raises it by 51 percentage points, from 51% to 102%, enabling the full one EV per home charging profile within the modeled line thermal limits. This coordinated flexibility therefore provides substantial additional EV demand headroom before thermal reinforcement is required. It also highlights the complementary value of DER support and topology reconfiguration in relieving feeder congestion under increasing EV demand.

TABLE III. MAXIMUM EV DEMAND SCALING UNDER ACTIVE POWER THERMAL FEASIBILITY CONSTRAINTS

| Configuration | Maximum feasible EV demand scaling |
|---|---|
| SDN | **51%** |
| SDN-DER | **68%** |
| SDNTR | **85%** |
| SDNTR-DER | **102%*** |

*100% corresponds to the one EV per residential home full adoption charging profile. Values above 100% represent uniform scaling of charging demand relative to this profile; therefore, 102% denotes 1.02 times the full adoption charging demand and does not indicate EV ownership above 100%.

Under the high demand profile, the boundaries decrease to 47.8%, 52.5%, 75.0%, and 79.7% for SDN, SDN-DER, SDNTR, and SDNTR-DER, respectively, while preserving the same configuration ranking. Coordinated DNTR and DER operation provides 31.9 percentage points of additional headroom over SDN, compared with 27.2 percentage points from DNTR alone.

## VII. Conclusion

This paper presents a data driven active power thermal feasibility assessment of EV integration with DERs and DNTR on a modified IEEE 33 bus system. The maximum feasible EV demand levels are 51%, 68%, 85%, and 102% for SDN, SDN DER, SDNTR, and SDNTR-DER, respectively. Under the high demand profile, these decrease to 47.8%, 52.5%, 75.0%, and 79.7%, while preserving the same configuration ranking. Coordinated DNTR and DER operation increases the expected profile boundary by 51 percentage points relative to SDN, enabling the full one EV per home charging profile within the modeled line thermal limits. Under the high demand profile, coordinated operation still provides 31.9 percentage points of additional feasibility headroom relative to SDN, showing that the benefit remains substantial under more demanding charging conditions. The results also indicate that DNTR provides the dominant feasibility improvement by redistributing power flows during peak charging periods, while DERs and BESS provide additional relief to constrained feeder corridors. The numerical limits are specific to the studied feeder and assumptions. Overall, the findings demonstrate the potential of coordinated DNTR and DER operation to increase EV demand accommodation and defer line thermal reinforcement. Because voltage, reactive power, losses, and transformer limits are not modeled, the reported values represent active power thermal feasibility boundaries rather than full AC hosting capacity limits.